\documentclass{svmult}

\usepackage{makeidx}     
\usepackage{graphicx}    
\usepackage{multicol}    

\makeindex             

\begin{document}

\title*{Investment strategy based on a company growth model
\vspace*{1.5cm}
}
\small\author{Takayuki Mizuno\inst{1},Shoko Kurihara\inst{1},
Misako Takayasu\inst{2}, and Hideki Takayasu\inst{3}}
\institute{Department of Physics, Faculty of Science and Engineering, Chuo 
University, 1-13-27 Kasuga, Bunkyo-ku, Tokyo 112-8551, Japan
\and Department of Complex Systems, Future University-Hakodate, 116-2 
Kameda-Nakano, Hakodate 041-0803, Japan
\and Sony Computer Science Laboratories, 3-14-13 Higashi-Gotanda, 
Shinagawa-ku, Tokyo, Japan}
%
%
\maketitle
\bigskip
\textbf{Summary. }We first estimate the average growth of a company's annual 
income and its variance by using both real company data and a numerical 
model which we already introduced a couple of years ago. Investment 
strategies expecting for income growth is evaluated based on the numerical 
model. Our numerical simulation suggests the possibility that an investment 
strategy focusing on the medium-sized companies gives the best asset growth 
with relatively low risk.

\vspace{1cm}


\vspace{0.3cm}

\section{Introduction}
\label{sec:1}
Studies on wealth distribution can be traced back more than 100 years. It is 
widely known that Pareto published the first report on the personal income 
distribution in 1897 [V. Pareto]. He showed that the probability density 
distribution of income follows a power law distribution in the high-income 
range. He claimed that a kind of social structure can be characterized by 
the power exponent because smaller value implies that the tendency of 
monopoly is stronger. 

By analyzing a huge database of company incomes, our group has been 
investigating statistical properties of time evolution of company's income 
[T. Mizuno(1), T. Mizuno(2)]. In those papers, we showed that prediction of 
the average income in the future is possible although the income of each 
company has violent fluctuations. A country needs to implement many policies 
of company's income because company's income is closely related to the tax. 
In the present paper we estimate the average growth rate of a company by 
using a numerical model that is derived from the real data and discuss the 
performance of investment strategy for companies.

\section{An empirical model for company's income growth}
\label{sec:2}
We have already introduced an empirical model describing the stochastic 
dynamics of company's income growth by the following equation [T. Mizuno(1), 
T. Mizuno(2)],

\vspace{0.2cm}
\begin{equation}
\label{eq1}
I_k (t + 1) = \alpha (t) \cdot b(t)^{\frac{\sigma (I)}{\sigma _0 }}I_k (t) + 
f(t)
\end{equation}
\vspace{0.2cm}

\noindent
where $I_k (t)$is the income of company k at the t-th year, $b(t)$ and 
$f(t)$ are random noises. The factor $\alpha (t)$ is a stochastic variable 
taking either 'P or --1. The probability densities of these functions can be 
estimated from the company data. The power exponent ${\sigma (I)} 
\mathord{\left/ {\vphantom {{\sigma (I)} {\sigma _0 }}} \right. 
\kern-\nulldelimiterspace} {\sigma _0 }$ characterizes the size dependence 
of company growth statistics. This model is simple but is confirmed to 
reproduce known empirical laws of company's income.

\section{The growth rate of company's income}
\label{sec:3}

For the purpose of discussing the investment strategy for income growth of 
American companies, we apply Eq.(1) with estimation the coefficient ($\alpha 
(t)$, $b(t)$, $f(t)$, ${\sigma (I)} \mathord{\left/ {\vphantom {{\sigma (I)} 
{\sigma _0 }}} \right. \kern-\nulldelimiterspace} {\sigma _0 })$ from a data 
set of income of American companies from 1989 to 1995 [T. Mizuno(2)][Moody's 
company data]. In order to find out an investment strategy we observe how a 
company of initial size $I(t=0)$ grows in the next 5 years. We calculate the 
averaged growth rate of the cumulative income, that is defined by the ratio 
of the cumulative income normalized by the initial income, $\Sigma 
I(t=5)/I(0)$, where $\Sigma I(t=5)=I(t=1)+ \cdots +I(t=5)$. In Fig.1 the numerical 
values of $\Sigma I(t=5)/I(0)$ estimated by Eq.(1) are compared with the 
values estimated directly from the real data of American companies (Ÿ). 
From this figure we notice that the expectation of the cumulative income 
$\Sigma I(t=5)$ is proportional to the initial income $I(0)$ for the 
companies whose initial income is larger than one million dollars/year. For 
the companies with initial incomes smaller than one million dollars/year the 
expectation value of cumulative income is nearly independent of the initial 
value of income. We can also estimate the income growth of a long term by 
using the numerical model. We show the case of $\Sigma I(t=15)$ by a line of 
USA{\_}S($t=15$) in Fig.1. Companies with initial incomes smaller than two 
million dollars/year also have the large average value of $\Sigma 
I(t=15)/I(0)$. These results imply that investing such smaller income 
companies will produce high growth on average. 

Although a small company's growth rate is large on average, its variance is 
also large. In Fig.2 we plot the probability densities of growth rates for 
the term of 5 years ($\Sigma I(t= 5)$) for three different initial income 
ranges. The probability density tails are very large for smaller companies 
as expected. In order to clarify the dependence on the initial income $I(0)$ 
for this variance, we show the standard deviation of the 

\vspace{1cm}
\begin{figure}[htbp]
\centerline{\includegraphics[width=7.2cm,height=4.5cm]{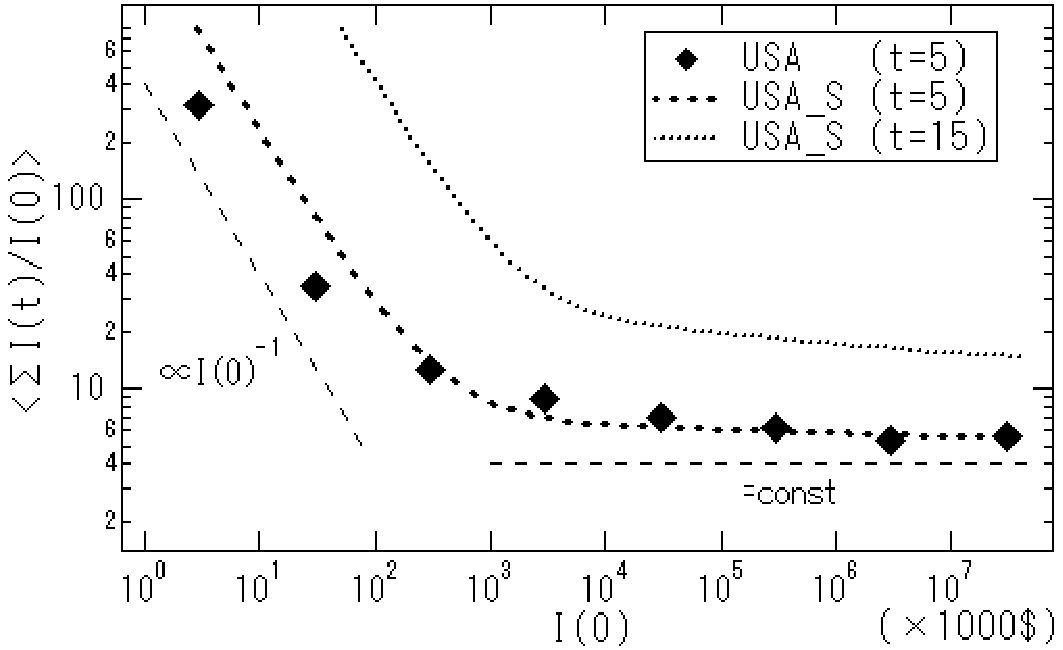}}
\textbf{Fig.1} Averaged growth rate as a function of the initial income. 
Plots of USA are real data for 5 years. Plots of USA{\_}S are results 
estimated by the numerical model.
\end{figure}

\begin{figure}[htbp]
\centerline{\includegraphics[width=7.1cm,height=4.5cm]{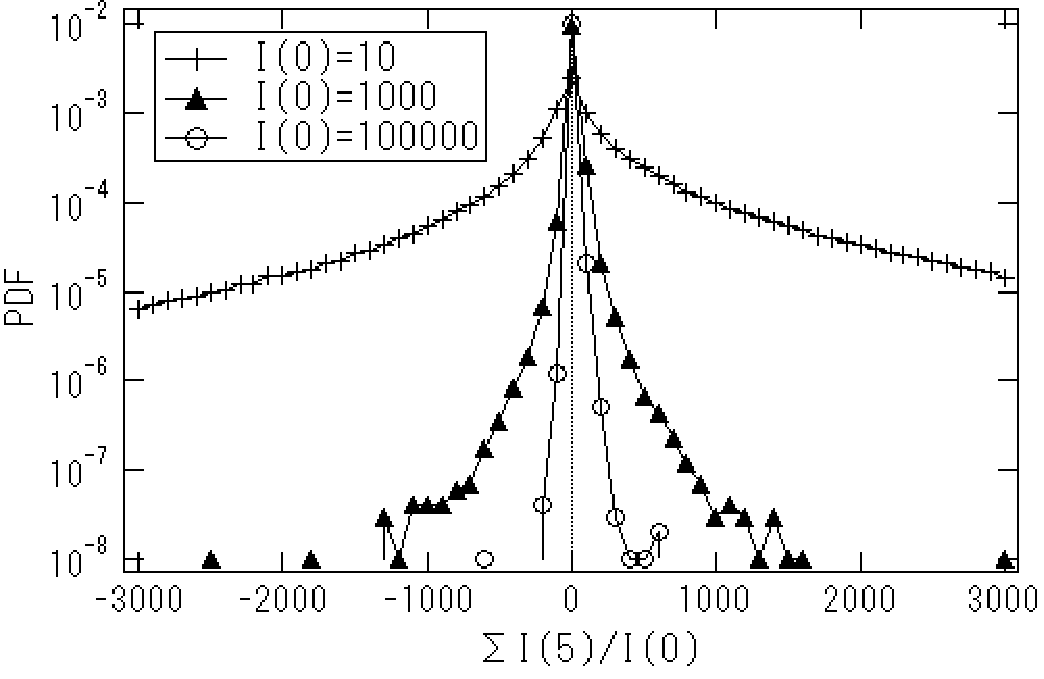}}
\center{
\textbf{Fig.2} The probability density of growth rate for 
the term of 5 years ($\Sigma I(t= 5)$).
}
\end{figure}

\begin{figure}[htbp]
\centerline{\includegraphics[width=7.3cm,height=4.5cm]{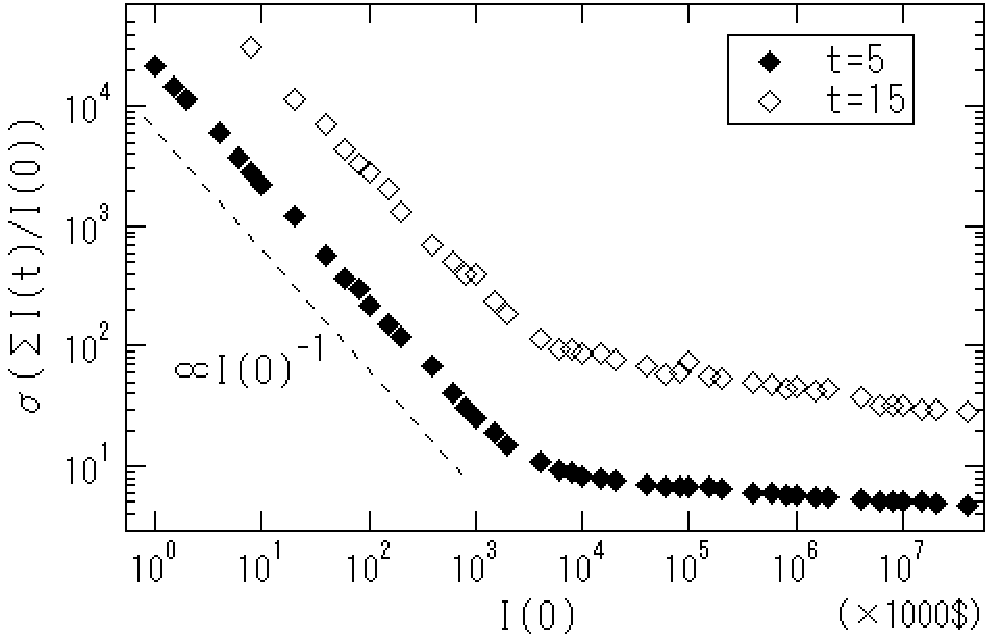}}
\center{
\textbf{Fig.3} The standard deviation of the growth rate.
}
\end{figure}

\noindent
probability density 
distribution of $\Sigma I(t=5)/I(0)$ in Fig.3. For companies with initial 
incomes smaller than one million dollars/year the standard deviation 
decreases rapidly in inverse proportion to initial income. For companies 
with initial incomes larger than one million dollars/year the standard 
deviation decreases slowly depending on initial income. Therefore, the risk 
caused by investment is small for large income companies.

\section{The efficient investment strategy}
\label{sec:4}
The relation between profits and risks is important when considering an 
efficient investment strategy. For the case of initial income 
$I(0)=10$ ($\times1000$ 
dollars) the distribution in Fig.2 is clearly asymmetric, namely, the 
average of return is positive. But there is a large possibility of taking 
very large loss because the fluctuations are so large. Namely, this case 
corresponds to the high-risk-high-return strategy, not a reasonable 
investment. In order to discuss a rational investment strategy with 
relatively low risk, we observe the relation between the average of $\Sigma 
I(t=5)/I(0)$ and the standard deviation. We define the investment efficiency 
$E(c, I(0))$ by the following equation,

\vspace{0.2cm}
\begin{equation}
\label{eq2}
E(c,I(0)) = \left\langle {{\sum {I(5)} } \mathord{\left/ {\vphantom {{\sum 
{I(5)} } {I(0)}}} \right. \kern-\nulldelimiterspace} {I(0)}} \right\rangle - 
c \cdot \sigma \left( {{\sum {I(5)} } \mathord{\left/ {\vphantom {{\sum 
{I(5)} } {I(0)}}} \right. \kern-\nulldelimiterspace} {I(0)}} \right),
\end{equation}
\vspace{0.2cm}

\noindent
where, $c$ is a constant showing the balances between profits and risks. We 
assume $c = $1/5`1/9 although this balance changes with contents of each 
investment. We show the relation between the investment efficiency $E$ and the 
initial income $I(0)$ in the case of $c = $1/8 in Fig.4. In this case the best 
reasonable investment targets are the companies of size around the peak 
point of the investment efficiency $E$, that is, the companies with income 
1,000,000`10,000,000 dollars/year for investment period of 5 years.

\begin{figure}[htbp]
\centerline{\includegraphics[width=7.3cm,height=4.5cm]{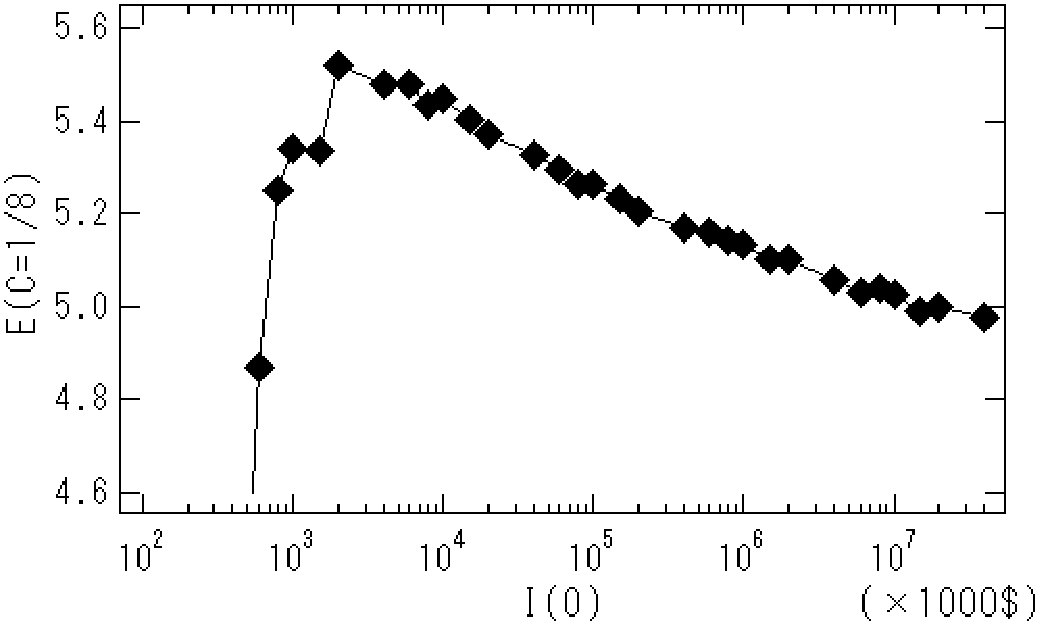}}
\center{
\textbf{Fig.4} The investment efficiency $E(c, I(0))$.
}
\end{figure}

\section{Discussion}
\label{sec:5}
We discussed the expected profits and risks for various investment 
strategies by using the simple model of company income and showed that 
investment to the medium-sized companies is efficient with relatively low 
risk. Although we showed results only for companies in USA, similar results 
hold also for other countries such as Japan and England. 

Turning our eyes to politics, Japanese government and banks tends to support 
large companies, however, judging from our results such investments can be 
categorized as too much low-risk-low-return strategy. Financial support to 
medium-sized companies will contribute to recovery of economical growth.

The correlation between the income and other indices (asset, etc.) [H. 
Takayasu] and the dynamics of other indices [M. H. R. Stanley] have been 
reported. We are going to use other indices of wealth in addition to income 
for the investment strategy in the future works. We expect that the 
investment strategy of high accuracy can be obtained by using these results.

\bigskip
\noindent
\textbf{Acknowledgement}

\medskip
\noindent
We would like to thank Prof. Tohru Nakano and Prof. Mitsuo Kono for stimulus 
discussions.

\bigskip
\noindent
\textbf{References}

\medskip
\noindent
V. Pareto, Le Cours d'Economie Politique, Macmillan, London, (1897).

\noindent
T. Mizuno, M. Katori, H. Takayasu and M. Takayasu, in \textit{Empirical Science of Financial Fluctuations} -- \textit{The Advent of Econophysics}, (Springer 
Verlag, Tokyo, 2002) 321-330.

\noindent
T. Mizuno, M. Takayasu, H. Takayasu, preparing for publication.

\noindent
Moody's company data 1989-1995. Big companies in U.S.A more than 10,000 
companies.

\noindent
H. Takayasu and M. Takayasu, \textit{Econophysics} -- \textit{Toward Scientific Reconstruction of Economy} (in Japanese, Nikkei, Tokyo, 2001).

\noindent
M. H. R. Stanley, et al, \textit{Nature} 379 (1996) 804-806.

\printindex
\end{document}